\documentclass{iau_FM}
\usepackage{graphicx}
\newcommand{\kepler}{{\it Kepler}}
\newcommand{\setone}{{\tt Set 1}}
\newcommand{\settwo}{{\tt Set 2}}

\title[Planetesimal Scattering of Near-Resonant Planets] 
{Period Ratio Distribution of Near-Resonant Planets Indicates Planetesimal Scattering}

\author[Chatterjee et\ al.]   
{Sourav Chatterjee$^1$, 
Seth O.~Krantzler$^1$
 \and Eric B. Ford$^{2,3}$}

\affiliation{$^1$Center for Interdisciplinary Exploration and Research in Astrophysics\\
Northwestern University\\2145 Sheridan Road, Evanston, IL 60208, USA \\ 
email: {\tt sourav.chatterjee@northwestern.edu} \\[\affilskip]
$^2$ Department of Astronomy \& Astrophysics \\
The Pennsylvania State University\\
 525 Davey Laboratory, University Park, PA 16802, USA \\[\affilskip]
$^3$ Center for Exoplanets and Habitable Worlds\\ 
The Pennsylvania State University\\ 
525 Davey Laboratory, University Park, PA, 16802, USA
}

\pubyear{2015}
\jname{Astronomy in Focus, Volume 1} 
\editors{Piero Benvenuti, ed.}
\begin{document}

\maketitle

\begin{abstract}
An intriguing trend among {\it Kepler}'s multi-planet systems is an overabundance 
of planet pairs with period ratios just wide of mean motion resonances (MMR) 
and a dearth of systems just narrow of them. In a recently published paper 
\cite{CF15} has proposed that gas-disk migration
traps planets in a MMR. After gas dispersal, orbits of these trapped planets are 
altered through interaction with a residual planetesimal disk. They found that for 
massive enough disks planet-planetesimal disk interactions can break resonances 
and naturally create moderate to large positive offsets from the initial period ratio for large 
ranges of planetesimal disk and planet properties. Divergence from resonance 
only happens if the mass of planetesimals that interact with the planets is at least a few percent of the total 
planet 
mass. This threshold, above which resonances are broken and the offset from resonances can grow, 
naturally explains why the asymmetric large offsets were not seen in more massive planet pairs found via past radial velocity 
surveys. In this article we will 
highlight some of the key findings of CF15. 
In addition, we report preliminary results from an extension of this study, that investigates 
the effects of planet-planetesimal disk interactions on initially non-resonant 
planet pairs. We find that planetesimal scattering typically increases period ratios of non-resonant 
planets. If the initial period ratios are below and in proximity of a resonance, under certain 
conditions, this increment in period ratios can create a deficit of systems with period ratios 
just below the exact integer corresponding to the MMR and an excess just above. From an initially uniform 
distribution of period ratios just below a 2:1 MMR, planetesimal interactions can create an asymmetric 
distribution across this MMR similar to what is observed for the \kepler\ planet pairs. 
\end{abstract}

\firstsection 
\section{Introduction}
\label{s:intro}
\begin{figure}[h]
\begin{center}
 \includegraphics[width=0.9\linewidth]{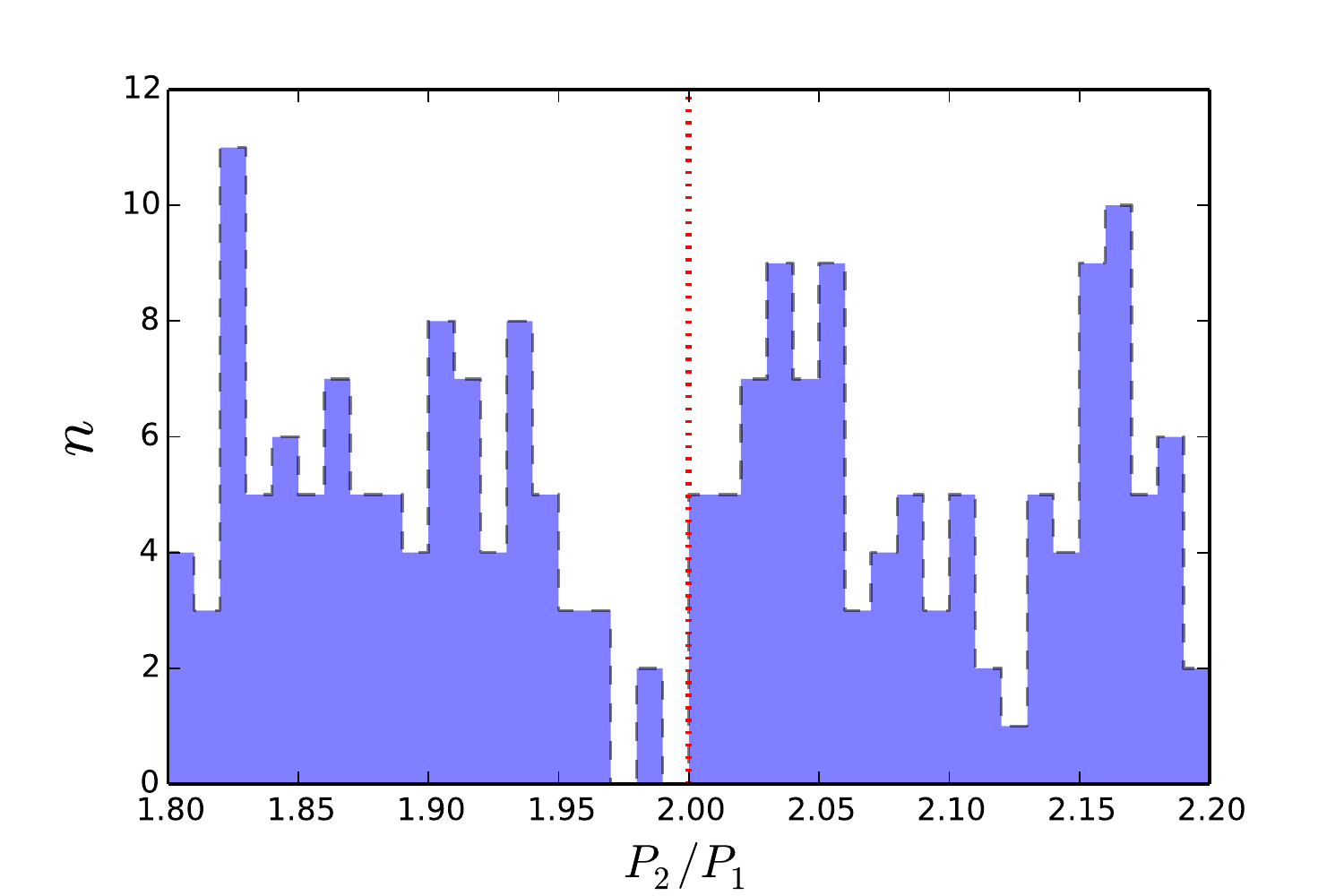} 
 \caption{Period ratio distribution of adjacent planet pairs discovered by \kepler\ close to the period ratio 
 expected for a 2:1 MMR. There is a dearth of systems with period ratios just narrow of the resonance and 
 an excess of systems just wide of the resonance. The vertical (red-dotted) line shows the exact position 
 of a period ratio of 2. \kepler\ data was extracted from NASA's exoplanet archive.  
 }
   \label{fig:obs}
\end{center}
\end{figure}

NASA's \kepler\ mission has revolutionized our understanding of planetary systems, 
their occurrence rate, multiplicity and physical properties. One trend apparent among this 
new class of small planets was a-priori quite unexpected from traditional theories; there is a 
statistically significant excess of planet pairs with period ratios slightly wide of first 
order mean motion resonances such as 2:1 and 3:2, and a dearth of them just narrow 
of these resonances (\cite{Lissauer2011, Fabrycky2014, Steffen2015}; Figure\ \ref{fig:obs}). 
Interestingly, this trend is absent in planets that were previously discovered 
via radial velocity (RV) surveys (e.g., \cite{Butler2006}). Smooth gas-disk migration can trap planets in 
MMRs. However, such resonant planets are expected to have period ratios with very small offsets from the 
integer ratio corresponding to the MMR, $\epsilon\equiv P_2/P_1 - (j+1)/j \sim \pm 10^{-3}$. 
Indeed, adjacent planet pairs discovered via past RV surveys show period ratio distribution with a distinct 
excess at the expected period ratio for the 2:1 MMR with very small 
$\epsilon$, consistent with the expectations from smooth gas-disk driven migration 
(\cite{LeePeale2002, Butler2006, Armitage2013}). In contrast, the near-resonant \kepler\ planet pairs 
are likely not in actual resonance (\cite{Veras2012}). Nevertheless, the overall close to uniform distribution 
away from resonance, and the mysterious asymmetric abundance 
across MMRs, such as 2:1 and 3:2, indicate that the \kepler\ planet pairs somehow knew about these resonances. 
However, some other process has driven them wide of the resonance and created this asymmetry. 
Planet-planet scattering 
can break resonances, however, they bring dramatic changes in the planetary orbits, often making 
them highly eccentric, which is inconsistent with the multi-transiting architecture of the \kepler\ systems 
(e.g., \cite{FR96, Chatterjee2008}). 
The large ($\epsilon\sim10\%$) positive offsets 
in the near-resonant planet pairs observed by \kepler\ thus has generated a lot of interest. 

It is generally 
believed that these planets were initially trapped in a MMR. Subsequently, some dissipative process 
drove them wide of their initial period ratios. The most likely dissipative mechanism responsible for 
the observed trend is still a matter of debate. The proposed dissipative mechanisms include 
dissipation from tide (\cite{LW2012, BM2013, DL2014}), turbulence in protoplanetary disk (\cite{Rein2012}), and 
scattering with a planetesimal disk (\cite{Moore2013}; CF15). Although, the most well studied, the 
tidal dissipation mechanism is also the most debated. \cite{LFL2013, Silburt2015} argue that even under generous 
assumptions, the large observed positive $\epsilon$ for most \kepler\ planet pairs near a MMR 
cannot be explained by tides alone. 
It was also suggested that in-place mass growth of a planet via planetesimal accretion can 
lead to formation of an over density of particles just wide of a MMR (\cite{Petrovich2013}). 
However, planetesimal accretion typically lead to migration of the planet making the in situ growth assumption 
questionable. It was also suggested that the 
observed period ratio distribution may be explained due to overstable libration of the \kepler\ planets 
due to gas-disk driven migration coupled with eccentricity damping (\cite{GS2014}). 
However, \cite{Hands2014, Deck2015} present an opposing view. 

In this article, we will focus on planet-planetesimal disk interactions as the mechanism for the observed 
asymmetric period ratio distribution across 2:1. Planet-planetesimal disk interactions have been well studied 
in other contexts, especially for the outer Solar system, and is generally believed to be a natural consequence 
of the core-accretion paradigm of planet formation (e.g., \cite{FernandezIp1984, Idaetal2000, Gomesetal2004, Kirshetal2009}). 
In \S\ref{s:numeric} we will 
briefly describe our numerical setup. In \S\ref{s:results} we will highlight our key results. First, we will 
highlight the findings of CF15 with some additional details. We will also present results from new simulations 
involving interactions of initially non-resonant planets with a planetesimal disk. Finally, in \S\ref{s:discuss} 
we will summarize our results and discuss the implications. 

\section{Numerical Setup}
\label{s:numeric}
The simulations presented in this article are from two distinct sets. One investigates the effects of 
planetesimal scattering on the orbits of two planets initially trapped in a 2:1 MMR (CF15). We call this \setone. 
The other investigates the effects of planetesimal scattering on planet orbits that are not initially trapped in 
2:1 MMR but, are just narrow of the MMR. We call this \settwo.

The details of the numerical setup for \setone\ are described in CF15. However, for completeness, we 
will briefly describe the key aspects. In general, the physical picture we have in mind is that while a gas disk 
is present, gas-disk interactions may trap two planets into 2:1 MMR. Once the gas disk is depleted, the resonant 
planets can freely interact with a residual planetesimal disk. We are interested in the effects 
of the latter interactions. Thus, we are interested in a system that initially was dissipative and transitions into 
a $N$-body. Ideally, planets, planetesimals, and a gas disk should be modeled together with all physics included, 
however, this full problem is computationally impractical. Hence, we generate plausible initial conditions for the stage of 
planet-planetesimal disk interactions in two steps. First, we use an analytic $\dot{a}$ and $\dot{e}$ prescription 
to trap two planets in 2:1 MMR (\cite{LeePeale2002}). Second, we create planetesimal orbits consistent with the 
presence of the planets in the following way. The structure of the residual planetesimal disk after gas-disk 
depletion is uncertain. Nevertheless, we use planetesimal disk profiles described by $d\Sigma/da \propto a^\alpha$, where 
$\Sigma$ and $a$ are the surface density and distance from the star for the planetesimals, respectively. 
At the epoch of gas disk depletion, the planetesimal disk profile would not remain a simple power-law. 
Instead, the planets would alter the planetesimal disk densities near them by 
dynamically scattering or accreting some of the nearby planetesimals that are on orbits unstable 
even with the stabilization provided by dissipation from a gas disk (e.g., \cite{Matsumura2010}). 
To imitate this effect we embed the resonant planets in a planetesimal disk with a power-law profile given by 
$d\Sigma/da \propto a^{\alpha}$. We treat all planetesimals as test particles. 
We let the planets alter the disk for at least $\sim 10^2$ orbits of the outer planet. We collect the properties 
of the planetesimals that survive this phase and create a database of allowed planetesimal orbits for each planet pair. 
We call this the clean-up stage. 
We randomly choose $2\times10^3$ orbits from 
this database, assign masses to the planetesimals according to the assumed planetesimal disk mass 
to planet mass ratio. 
We evolve the resonant planets with the planetesimals until the period ratio of the planets' orbits stop 
changing ($\sim 10^5$ years). CF15 has varied the planetesimal 
disk profile by changing the power-law index $\alpha$ between -2.5 to 3, planet-planet mass ratios $m_1/m_2$ between 
0.1 to 10, and the planetesimal disk to planet mass ratio ($m_d/(m_1+m_2)\equiv m_d/m_p$) between 0.1 to 1.5.  

We have started investigating the effects of a residual planetesimal disk on initially non-resonant planets. 
In this set, \settwo, we closely follow the prescriptions of CF15 summarized above. However, we choose planet pairs 
with initial orbits such that the initial period ratio is between 1.8 and 2, just narrow of the 2:1 MMR. The inner planet's  
semimajor axis is kept at $a_1=0.1\ \rm{AU}$, a typical value for the \kepler\ planets. 
We generate 50 systems such that the period ratio $P_2/P_1$ is between 
1.8 and 2. The eccentricities of the planets' orbits are drawn from a Rayleigh 
distribution with scale 0.005 (e.g. \cite{Hadden2014}). The orbital inclinations ($I$) are drawn uniformly in $\cos I$ with 
$I$ between $-0.1$ and $0.1^{\circ}$. Planets have equal mass and are equal to the mass of Neptune ($M_N$) and 
have Neptune-like densities.  
For each pair of planetary orbits created 
this way, we generate 4 random realizations varying the phase angles in their full range. 
The initial planetesimal disk profile is given by $d\Sigma/da \propto a^{-3/2}$. The planetesimal disk edges 
are set at orbits with period $P_1/3$ and $3P_2$. For each planet pair, 
we perform the initial cleaning up of planetesimals exactly the same way as prescribed by CF15. For each case, a 
database of orbits is generated. We randomly select $2\times10^3$ orbits and give each planetesimal 
a mass $m_{pl}=1\times10^{-3}\ M_N$, such that $m_d/m_p = 1$. We integrate 
the planet-pair and planetesimals using the Bulirsch-Stoer integrator included in Mercury 
(\cite{Chambers1999}). We stop 
our integrations at $2\times10^4$ year, equivalent to $\approx 6\times10^5$ of the inner planet's initial orbital period. 
We confirm that the majority of the 
planetesimal interactions happen much earlier than our chosen integration stopping time.  

\begin{figure}[t]
\begin{center}
 \includegraphics[width=0.9\linewidth]{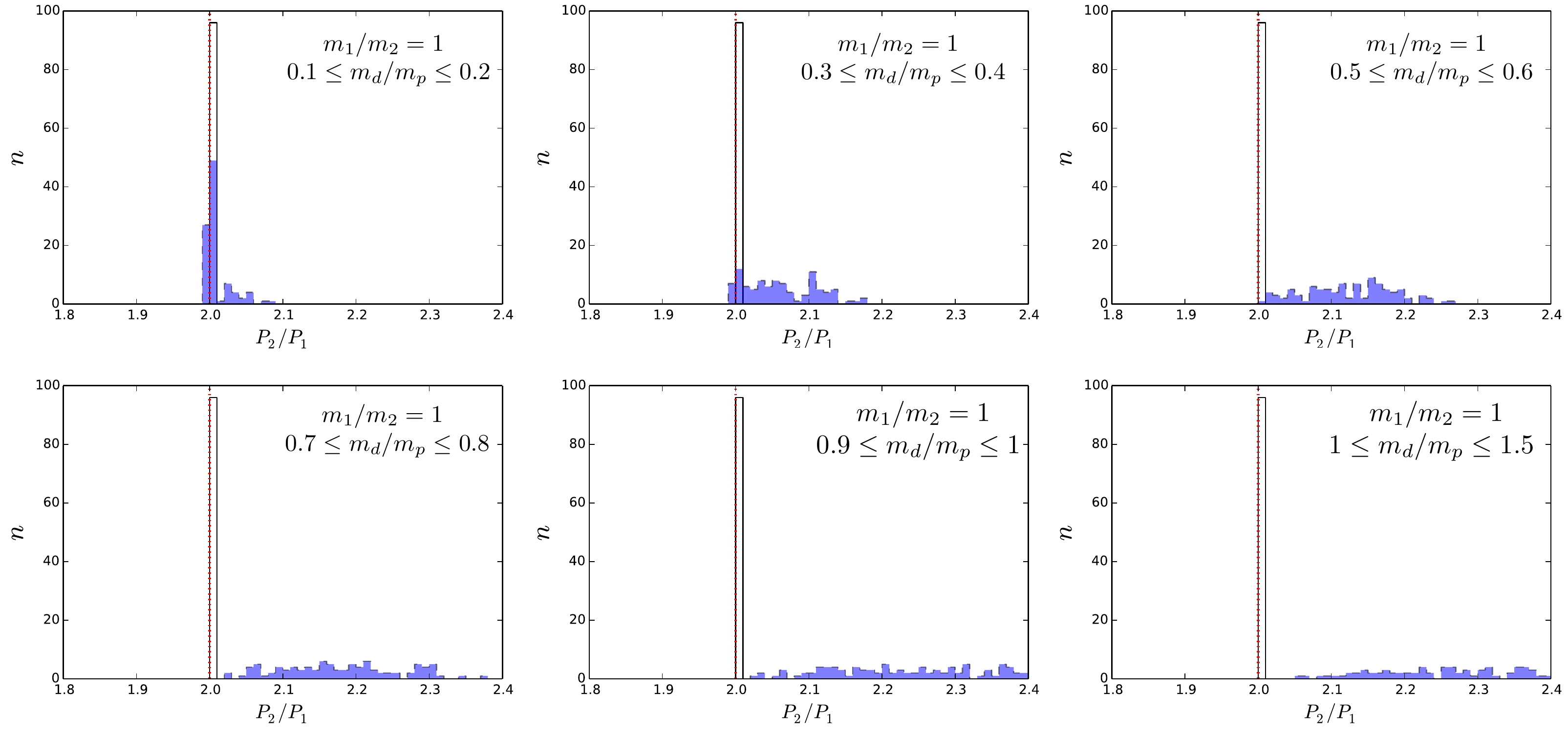} 
 \caption{The initial (black) and final (blue filled) period ratio distributions for models with equal mass planets 
 and planetesimal disk profile given by $\alpha=-3/2$ (a subset of models 
 presented in CF15). Each panel shows results from models with a different $m_d/m_p$ value, listed in each panel. 
 In all cases, the planet pairs are initially trapped in a 
 2:1 MMR. As a result, the initial period ratios are always close to 2. Depending on $m_d/m_p$ the final period ratio 
 distribution changes. For $m_d/m_p<0.3$ resonance is not broken for most planet pairs. As $m_d/m_p$ increases, so 
 fraction of systems for which resonance is broken and the highest $\epsilon$ attained, both increase. 
  }
   \label{fig:reso}
\end{center}
\end{figure}
\section{Results}
\label{s:results}
CF15 has investigated the effects of planetesimal interactions on the orbits of planet pairs initially trapped in a 2:1 MMR. The key 
results of CF15 are as follows. If the total mass of planetesimals that had strong interactions with the planets 
is high enough to break resonance, then 
planet-planetesimal interactions naturally increase the period ratio. The final offset from the MMR depends 
strongly on the ratio of the total mass of planetesimals that interacted with the planets and the planet mass. 
When resonance is broken, offset $\epsilon$ can have large positive values. If the resonance 
is not broken due to insufficient mass in nearby planetesimals, 
$\epsilon$ remains small ($\sim 10^{-3}$) and can have both positive or negative values. As a result, 
it is easier to break resonance and create large positive $\epsilon$ for low-mass planet pairs typical of those 
discovered by \kepler, compared to the much higher mass planet pairs discovered via past RV surveys. 
Figure \ref{fig:reso} shows the initial and final period ratio distribution from a subset of simulations presented 
in CF15. As the ratio $m_d/m_p$ increases, so does the fraction of systems where the resonance is broken 
and also the value of the highest $\epsilon$ the planet pairs can attain. 

\begin{figure}[t]
\begin{center}
 \includegraphics[width=0.9\linewidth]{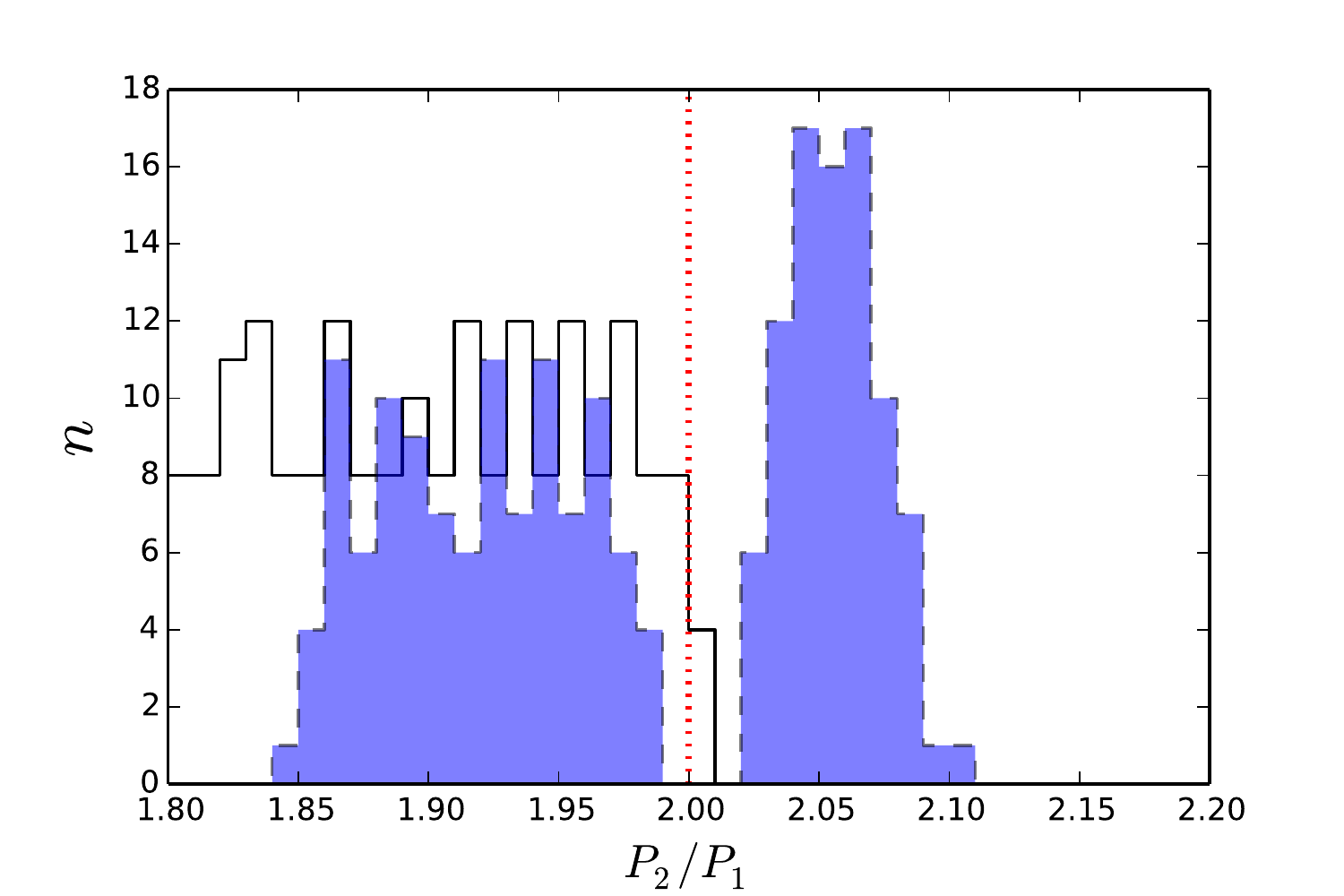} 
 \caption{Initial (black) and final (blue filled) period ratio distributions for the set of simulations including 
 initially non-resonant planets and a planetesimal disk (Section\ \ref{s:numeric}). Initially $m_1/m_2=1$,  
 $\alpha=-3/2$, and $m_d/m_p=1$. As a result of planetesimal interactions the period ratios 
 generally increase. As the pairs approach 2:1 MMR from the inside, the resonance is skipped and the period ratio 
 increases to a value $>2$. Thus, a dearth of planet pairs with period ratios just smaller than 2 and an excess of pairs 
 with period ratios just higher than 2 are created. }
   \label{fig:noreso}
\end{center}
\end{figure}

CF15 results suggest that planetesimal interactions with resonant planet pairs can naturally redistribute 
these planet pairs wide of the initial resonance. However, CF15 do not directly 
address the dearth of planet pairs with period ratios just 
narrow of 2:1. We have started a systematic study of the effects of planetesimal scattering on planet pairs that 
are initially not in resonance, rather has period ratios slightly smaller than 2 (\settwo). This is equivalent to a scenario 
where planet pairs do not go through significant migration in a gas disk. Thus, period ratios below and up to 2 
are populated uniformly before planetesimal scattering can take place. Figure\ \ref{fig:noreso} shows 
the initial and final period ratio distributions for a large ($2\times10^2$) set of simulations, each involving two planets 
and $2\times10^{3}$ planetesimals (\S\ref{s:numeric}). Initially, the planet pairs have period ratios uniformly 
spaced between 1.8 and 2. Interactions with planetesimals from the disk generally increase the period ratios. 
Interestingly, as some of the planet pairs approach the 2:1 MMR via planetesimal driven divergent migration, 
they tend to skip the resonance and get deposited wide of the resonance. As a result, a 
dearth of systems is created with period ratios slightly below 2 and an excess of systems is created with period ratios 
slightly above 2. 

\begin{figure}[t]
\begin{center}
 \includegraphics[width=0.9\linewidth]{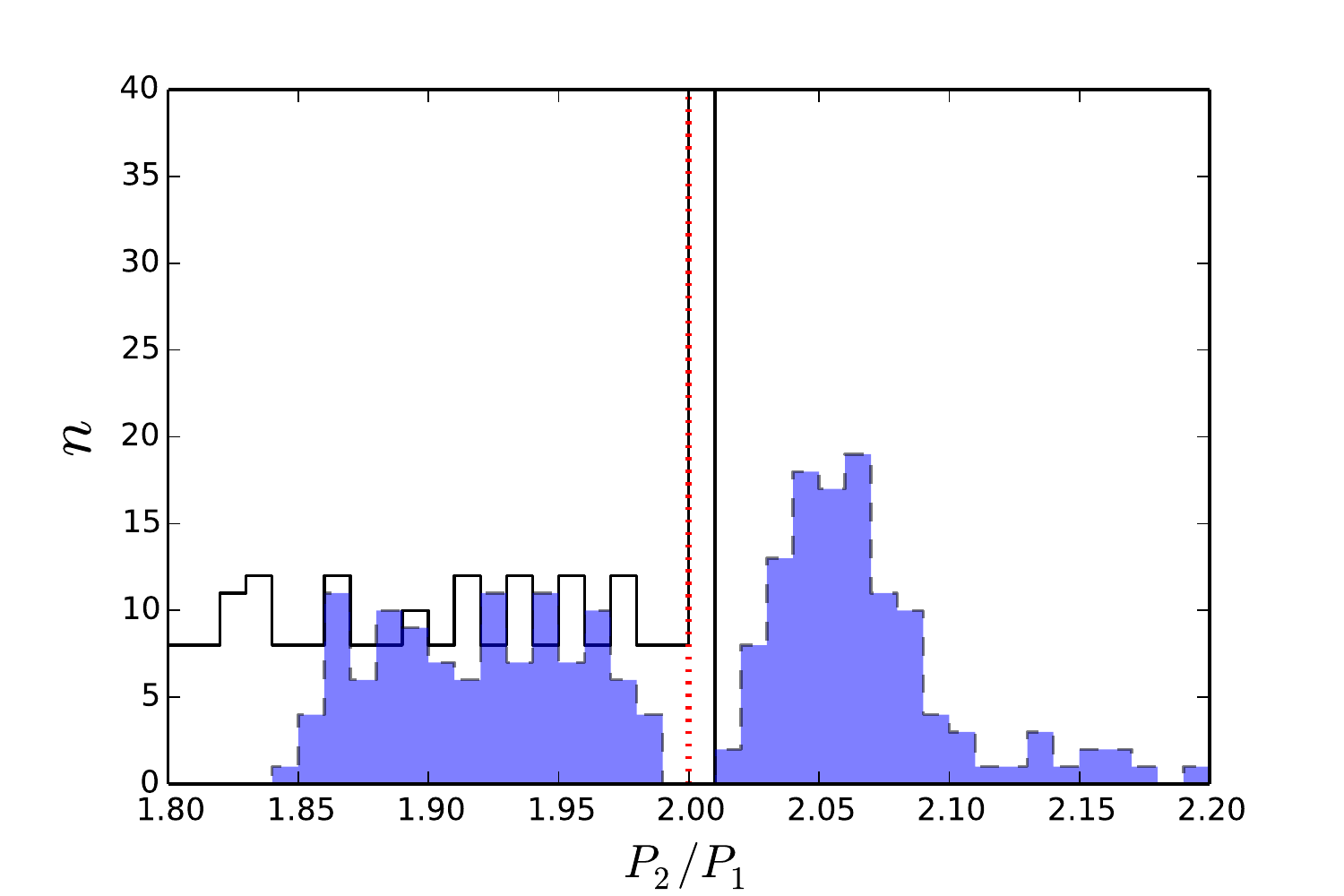} 
 \caption{The same as Figure\ \ref{fig:noreso} but combining models from CF15 with $m_1/m_2=1$, $\alpha=-3/2$, and 
 $m_d/m_p\geq0.3$ and those from \settwo\ (\S\ref{s:numeric}). }
   \label{fig:combined}
\end{center}
\end{figure}

In Figure\ \ref{fig:combined} we combine the models from CF15 with $m_1/m_2=1$, $\alpha=-3/2$ and $m_d/m_p\geq0.3$ 
with those from \settwo. Although the relative initial abundances between initially 
resonant and non-resonant planets is somewhat ad-hoc, the combined period ratio distribution is qualitatively very similar 
to what is observed across the 2:1 MMR for \kepler's adjacent planet pairs. 
Hence, immediately after gas disk dispersal, if planet pairs have period ratios distributed uniformly below 
2:1 with some excess of pairs at 2:1, and there is sufficient mass in nearby planetesimals in a residual planetesimal disk, 
then after planetesimal interactions the final period ratio distribution will, at least 
qualitatively, be very similar to what is observed of the \kepler\ planet pairs near 2:1 (Figure\ \ref{fig:obs}). 

\section{Discussion}
\label{s:discuss}
In this article we summarize the key results of CF15. In addition, we present preliminary results from a new set of simulations 
with initially non-resonant planets and a planetesimal disk. Our results suggest that planetesimal 
scattering may be responsible for both the excess of planet pairs just wide of the 2:1 MMR and the dearth of 
systems just narrow of it. 

We made several simplifying assumptions in this article. For example, in the new set of simulations with non-resonant 
planet pairs, we assume that the initial period ratios are distributed uniformly below 2 and all the way up to 2. This of course is 
not necessarily true in reality. Sufficiently far away from a MMR the period ratios of \kepler's planet pairs appear to 
be random. However, they may not have been so immediately after gas disk dispersal. 
In reality, there may have been a dearth of systems just narrow of a MMR simply because some systems,  
while migrating within a gas disk, got trapped into the MMR. In that sense, our assumption of uniform period ratio 
distribution below the 2:1 MMR is the most 
conservative one. Even if after gas disk dispersal the period ratios are uniformly distributed narrow of the 
2:1 MMR, planetesimal scattering from a sufficiently massive disk can create a dearth of systems with period ratios 
just below 2. These systems, in turn, pile up with period ratios slightly above 2. Encouraged by our preliminary results,  
we are exploring this problem more thoroughly by covering a larger parameter space and obtaining a more detailed understanding 
of the evolution of the planet pairs as they cross the 2:1 MMR from inside out via planetesimal driven migration.  

\acknowledgements
This research has made use of the NASA Exoplanet Archive, which is operated by the California 
Institute of Technology, under contract with the National Aeronautics and Space Administration 
under the Exoplanet Exploration Program.
We thank Frederic A. Rasio for helpful comments and discussions. The CF15 simulations were done 
using the High Performance computing (HPC) center at University of Florida. The simulations involving initially 
non-resonant planets were done using the HPC Quest at CIERA, Northwestern University. SC was partially 
supported by NASA grant NNX12AI86G. SOK thankfully acknowledges NASA's summer research grant. 
EBF was supported in part by NASA Kepler Participating Scientist Program award NNX14AN76G. 
The Center for Exoplanets and Habitable Worlds is supported by the Pennsylvania State University, 
the Eberly College of Science, and the Pennsylvania Space Grant Consortium.

\vspace{10 mm}
{\bf Questions and Comments:}

{QUESTION:}  In your simulations you have ignored planetary migration. If the planets migrate in a gas disk 
it will clear a larger gap in the planetesimal disk. As a result, after gas dispersal there won't be many planetesimals 
near the planets and interactions will be rare. Can you comment on that? 

{CHATTERJEE:} Indeed, planets would 
migrate in a gas disk. So would the planetesimals. In fact, the planetesimals are expected to migrate faster than the planets. As a 
result, following gas dispersal, the surface density of planetesimals may be enhanced exterior to the planets. 
That is why we considered a wide variety of planetesimal surface density profiles, including 
those for which the surface density increases with the distance from the star. 

Of course, the ideal way to study this problem is to simulate gas disks, planets, and planetesimals 
all together including all relevant physical effects.  Such simulations are unfortunately numerically 
impractical. Hence we are forced to adopt a scheme that mimics the expected configuration of a 
system that was dissipative initially and then became pure $N$-body.  In particular, we constructed 
our initial conditions in two steps, first trapping planets in resonance, and then removing planetesimals 
that would be unstable on short timescales. We believe that this could mimic the configuration of 
planets and planetesimal disks at the epoch of gas dispersal, at least qualitatively. A larger gap in 
planetesimals would slow down subsequent interactions with the planets, but as long as there are 
enough planetesimals to interact with the planets, the planetary orbits will diverge leading to a ratio 
of orbital periods greater than that of the exact resonance. Given that the general outcome of 
planet-planetesimal interactions is unchanged for the wide range in explored
planetesimal disk properties, we believe that the details of planetesimal disk structures are unlikely 
to change our results qualitatively.  

\end{document}